\documentclass[epjST]{svjour}
\usepackage{graphics}
\begin{document}
\title{Nonequilibrium properties of an atomic quantum dot coupled to a Bose-Einstein condensate}

\author{Roberta Citro\inst{1}\fnmsep\thanks{\email{citro@sa.infn.it}} \and Adele Naddeo\inst{2}\fnmsep\thanks{\email{naddeo@sa.infn.it}} \and Francesco Romeo\inst{1}\fnmsep\thanks{\email{fromeo@sa.infn.it}} }
\institute{Dipartimento di Fisica "E. R. Caianiello",
Universit\'{a} degli Studi di Salerno and CNR-SPIN, Unit\'{a} Operativa di Salerno, Via Ponte Don Melillo, 84084 Fisciano (SA),
Italy \and Dipartimento di Fisica "E. R. Caianiello",
Universit\'{a} degli Studi di Salerno, and CNISM, Unit\'{a} di
Ricerca di Salerno, Via Ponte Don Melillo, 84084 Fisciano (SA),
Italy}
\abstract{
We study nonequilibrium properties of an atomic quantum dot (AQD)
coupled to a Bose-Einstein condensate (BEC) within Keldysh-Green's
function formalism when the AQD level is varied harmonically in
time.
Nonequilibrium features in the AQD energy absorption spectrum are the side peaks that
develop as an effect of photon absorption and emission. We show that atoms
can be efficiently transferred from the BEC into the AQD
for the parameter regime of current experiments with cold atoms.
}
\maketitle
\section{Introduction}
\label{intro}
The tremendous experimental progress in the field of ultracold atoms has
made possible with unprecedent control the realization and design of
many-body atomic physics by means of very stable optical lattices and by the
use of Feshbach resonances\cite{cd1}\cite{cd2}. That has triggered the study
of such systems for quantum information processing purposes \cite{cd3} and
as quantum simulators of condensed matter Hamiltonians \cite{cd4}. In this
context the experimental development of atom chips \cite{cd5} opened new
perspectives towards atomic mesoscopic physics. Indeed the possibility to
build up atomic waveguides above microfabricated surfaces allows one to
study a variety of phenomena such as the interplay between interaction and
transport through point contacts \cite{cd6}, the dynamics of soliton-like
structures in waveguides \cite{cd7} and the coherent flow of a Bose Einstein
condensate (BEC) through a double barrier potential realized as a quantum
dot in a magnetic waveguide \cite{cd8}. \newline
Another route towards trapping and manipulation of single atoms \cite{cd9}
has been recently proposed, which relies on a focused laser beam
superimposed to a trap holding an atomic Bose Einstein condensate; that
leads to the formation of an atomic quantum dot (AQD) \cite{aqd1}, i.e. a
single atom in a tight trap, which is coupled to a superfluid reservoir via
laser transitions. In particular, in the case of a one dimensional
superfluid reservoir, the system maps onto a spin-boson model with ohmic
coupling, which exhibits a dissipative quantum phase transition\cite
{spinboson} and provides a spectroscopic tool to engineer open quantum
systems and study their irreversible dynamics \cite{aqd2}. Such a system can
also be employed as a probe of BEC phase fluctuations via dephasing
measurements of the internal states of the AQD, as discussed in %
\cite{aqd3}. More recently an AQD coherently coupled by optical
transitions to two BEC reservoirs has also been considered together with a
Josephson tunneling between the two BECs \cite{posa1}. Finally, a bosonic
single-impurity Anderson model (AM) has been recently numerically
investigated\cite{bulla1} in order to understand the local dynamics of an
AQD coupled to a BEC system, accounting for the entanglement and the
decoherence of the macroscopic condensate. In general, the strong collisional interaction in the AQD can
locally break a BEC state to bring up the excitations of normal
particles inside the dot and thus a non-equilibrium particle
transport can be realized, while the number of excited particles in
the QD gives an indirect probe of the coherence of the macroscopic
condensate.

In this work we discuss the realization of coherent
particle transport in an AQD resonantly coupled to a BEC via a
Raman transition with effective Rabi frequency and realized by
harmonic time modulation of the AQD level. We do consider a fully
quantum non-equilibrium situation within a Keldysh approach, not simply looking at the stationary limit, and
focus on the local spectral properties of the QD.

\section{Model and results}
\label{sec:1}
In our model system we consider cold bosonic atoms with two hyperfine ground
states $a$ and $b$. Atoms in state $a$, which form the reservoir, are
confined in a shallow trapping potential $V_{a}\left( \mathbf{x}\right) $
while atoms in state $b$ are localized in a tightly confining potential $%
V_{b}\left( \mathbf{x}\right) $.  The atoms are identical bosons
of mass $m$ and form a Bose-Einstein condensate in a state $a$,
i.e. a quantum reservoir. Furthermore states $a$ and $b$ are
coupled via a Raman transition with effective Rabi frequency
$\Omega $. In the collisional blockade limit for atoms $b$ only a
single atom is localized on
the dot. Atoms $a$ and $b$ are described by the field operators $\widehat{%
\Psi }_{a}$ and $\widehat{\Psi }_{b}$, with $\widehat{\Psi }_{b}\left(
\mathbf{x}\right) =\psi _{b}\left( \mathbf{x}\right) \widehat{b}$, where the
operator $\widehat{b}$ destroys an atom $b$ in the dot in the lowest
vibrational state with wave function $\psi _{b}\left( \mathbf{x}\right) $
and $\widehat{\Psi }_{a}\left( \mathbf{x}\right) \sim \widehat{\rho }%
_{a}\left( \mathbf{x}\right) ^{1/2}e^{-i\widehat{\phi }\left( \mathbf{x}%
\right) }$ in the long-wavelength limit, where $\hat{\rho}_a$ and $\hat{\phi}$ are the density and phase operator, respectively. The general many-body Hamiltonian of our system
is \cite{aqd1}:
\begin{equation}
H=H_{a}+H_{b}+H_{ab}  \label{m1}
\end{equation}
where
\begin{equation}
H_{a}=\frac{1}{2}\int d\mathbf{x}\left( \frac{\hbar ^{2}}{m}\rho _{s}\left|
\nabla \widehat{\phi }\right| ^{2}\left( \mathbf{x}\right) +\frac{mv_{s}^{2}%
}{\rho _{a}}\widehat{\Pi }^{2}\left( \mathbf{x}\right) \right) ,  \label{m2}
\end{equation}
and
\begin{eqnarray}
H_{b}+H_{ab}=\left( -\hbar \delta _{0}+g_{ab}\int d\mathbf{x}\left| \psi
_{b}\left( \mathbf{x}\right) \right| ^{2}\widehat{\rho }_{a}\left( \mathbf{x}%
\right) \right) \widehat{b}^{+}\widehat{b}+ \nonumber \\
\frac{U_{bb}}{2}\widehat{b}^{+}%
\widehat{b}^{+}\widehat{b}\widehat{b}+\hbar \Omega(t) \int
d\mathbf{x} \cos( \mathbf{k} \cdot \mathbf{x}) \left(
\widehat{\Psi }_{a}\left( \mathbf{x}\right) \widehat{\Psi
}_{b}^{\dagger }\left( \mathbf{x}\right) e^{i\omega_p t}
+h.c.\right), \label{m3}
\end{eqnarray}
where the first term in Eq. (\ref{m3}) contains the Raman detuning $\delta
_{0}$ and the collisional interactions between the dot and the reservoir, $%
g_{ab}=\frac{4\pi \hbar ^{2}a_{ab}}{m}$, with effective mass $%
m=m_{a}m_{b}/(m_{a}+m_{b})$ and $s$-wave scattering lengths $a_{ab}$;
 $\widehat{\rho }_{a}\left( \mathbf{x}\right) =\widehat{\Psi }%
_{a}^{+}\left( \mathbf{x}\right) \widehat{\Psi }_{a}\left(
\mathbf{x}\right) $ is the density operator for atoms $a$, which
can be expressed in terms of the density fluctuation operator
$\widehat{\Pi }$ (canonically conjugated to $\widehat{\phi }$) as
$\widehat{\rho }_{a}\left( \mathbf{x}\right) =\rho
_{a}+\widehat{\Pi }\left( \mathbf{x}\right)$; the second term
describes the interaction strength of two-body s-wave collision
$U_{bb}\propto g_{bb}$ between the dot atoms (which we do consider in
the collisional regime) while the third one is
the laser induced coupling between atoms $a$ and $b$ with Rabi frequency $%
\Omega $. We do consider a step-like pulse with frequency
$\omega_p$ and duration $\tau$. Since in the collisional blockade
limit $\omega_{gap} \tau \gg 1$ where $\omega_{gap}$ is the gap
between the single and two-atom, the dynamics describing the
coupling to all other bound states can be neglected and the laser
resonantly couples the condensate with the single atom ground
states of the dot with an effective strength
$V_{\mathbf{k}}=\hbar \Omega \int d%
\mathbf{x}\cos (\mathbf{kx})\psi _{a}(\mathbf{x})\psi
_{b}(\mathbf{x})$, i.e. the Fourier transform of the wavefunctions
overlap. At low enough temperatures, the reservoir Hamiltonian
(\ref{m2}) is that of a Bose superfluid with equilibrium density
$\rho _{a}$, superfluid density $\rho _{s}$ and low energy phonon
excitations of linear dispersion $\omega =v_{s}\left|
\mathbf{q}\right| $, $v_{s}$ being the sound velocity. Eq.
(\ref{m2}) can also be written as a bath of harmonic sound modes:
\begin{equation}
H_{a}=\hbar v_{s}\sum_{\mathbf{q}}\left| \mathbf{q}\right| b_{\mathbf{q}%
}^{+}b_{\mathbf{q}},  \label{m4}
\end{equation}
where $b_{\mathbf{q}}$ are standard phonon operators, defined by:
\begin{equation}
\begin{array}{c}
\widehat{\phi }\left( \mathbf{x}\right) =i\sum_{\mathbf{q}}\left| \frac{%
mv_{s}}{2\hbar \mathbf{q}V\rho _{a}}\right| ^{1/2}e^{i\mathbf{q\cdot x}%
}\left( b_{\mathbf{q}}-b_{-\mathbf{q}}^{+}\right)  \\
\widehat{\Pi }\left( \mathbf{x}\right) =\sum_{\mathbf{q}}\left| \frac{\hbar
\mathbf{q}\rho _{a}}{2mv_{s}V}\right| ^{1/2}e^{i\mathbf{q\cdot x}}\left( b_{%
\mathbf{q}}+b_{-\mathbf{q}}^{+}\right)
\end{array}
,  \label{modes1}
\end{equation}
$V$ being the system volume. The Hamiltonian (\ref{m1}) can be
reduced to that of a spin-boson model \cite{spinboson} in the collisional blockade limit (i.e. large strength
$U_{bb}$) when the internal state of the AQD is described by a
pseudospin $1/2$ with the spin-up and spin-down state
corresponding to single and no atom occupation respectively. Thus
the following replacements can be taken:
$\widehat{b}^{+}\widehat{b}\rightarrow \frac{\left( 1+\hat{\sigma}
_{z}\right) }{2}$ and $\widehat{b}^{+}\rightarrow \hat{\sigma
}_{+}$, where $\hat{\sigma }_{+,z}$ stand for spin operators.

In order to describe the density fluctuations regime and to
conserve the total number of bosons, the previous model can be
conveniently mapped onto a bosonic single-impurity Anderson model
(AM)\cite{bulla1}:
\begin{equation}
H=\varepsilon b^{+}b+\frac{U}{2}b^{+}b\left( b^{+}b-1\right) +\sum_{\mathbf{k%
}}\varepsilon _{\mathbf{k}}b_{\mathbf{k}}^{+}b_{\mathbf{k}}+\sum_{\mathbf{k}%
}V_{\mathbf{k}}(t)\left( b^{+}b_{\mathbf{k}}+b_{\mathbf{k}}^{+}b\right) .
\label{m7}
\end{equation}
where $b_{\mathbf{k}}$, $b_{\mathbf{k}}^{+}$ are annihilation and creation
operators of noninteracting bosons confined in a shallow potential, $%
\varepsilon $, $U$ are immediately recognized as a function of the detuning
and the collisional energies of (\ref{m3}), $\varepsilon _{\mathbf{k}}=v_{s}|%
\mathbf{k}|$ and the last term is the laser-induced hybridization
between particles in the AQD and the bosonic bath where
$V_{\mathbf{k}}$ has been given above. An estimate of the
Hamiltonian parameters is the following \cite{morigi1}: the
reservoir can be made of a condensate of $N\simeq 10^{3}$
$^{87}Rb$ atoms with density $n_{b}=3\cdot 10^{13}$ $atoms/cm^{3}$
in a harmonic trap (with trapping frequency $\nu _{b}=2\pi \times
100Hz$), while the tweezers trap for the AQD in the collisional
regime should have frequencies of the order of hundreds of $kHz$
and we take $\nu _{a}=2\pi \times 100kHz$ and a detuning $\delta
_{0}=284kHz$. The Raman coupling between the dot and the reservoir
for these parameters can be estimated of the order of $\Omega
\simeq 30kHz$.
\newline

Since we are interested in the nonequilibrium properties and
particle transport through the AQD, we do consider a situation
in which the AQD level is harmonically varied in time via the coupling to the radiation, i.e.
$\epsilon \rightarrow \epsilon (t)=\epsilon_0 +\epsilon_\omega
\cos \omega t$, and employ the real-time Keldysh formalism to
calculate the spectral properties of the AQD and particle
transport through it. Since in the following we are interested in the collisional blockade limit of the AQD, the dot levels $\epsilon_0$ and $\epsilon_\omega$ have to be considered not as bare ones, but as renormalized by the requirements of a single atom occupation.

The BEC energy absorption and the
particle current flowing through the AQD are given by:
\begin{eqnarray}
\left\langle \dot{E}_{\mathbf{k}}(t)\right\rangle  &=&-\frac{i}{\hbar }%
\left\langle [E_{\mathbf{k}},H(t)]\right\rangle,  \\
J\left( t\right)  &=&-\left\langle \frac{d{N}}{dt}\right\rangle =\frac{-i}{%
\hbar }\left\langle \left[ H,N\right] \right\rangle   \label{m12}
\end{eqnarray}
where $E_{\mathbf{k}}=\varepsilon _{\mathbf{k}}b_{\mathbf{k}}^{+}b_{\mathbf{k}}$ is the BEC Hamiltonian and $N=\sum_{%
\mathbf{k}}b_{\mathbf{k}}^{+}b_{\mathbf{k}}$. By computing the commutators, both the BEC energy absorption and the particle current can be expressed in
terms of the lesser Green's function $G_{0,\mathbf{k}}^{<}\left(
t,t^{^{\prime }}\right) \equiv -i\left\langle b_{\mathbf{k}}^{+}\left(
t^{^{\prime }}\right) b\left( t\right) \right\rangle $ as:
\begin{eqnarray}
\left\langle \dot{E}_{\mathbf{k}}(t)\right\rangle  &=& \varepsilon_{\mathbf{k}} J_{\mathbf{k}}, \nonumber \\
J\left( t\right)&=&  \sum_{\mathbf{k}} J_{\mathbf{k}}
\end{eqnarray}
where
\begin{eqnarray}
J_{\mathbf{k}}=-\frac{2}{\hbar }%
Re\left\{V_{\mathbf{k}}\left( t\right) G_{0,%
\mathbf{k}}^{<}\left( t,t\right) \right\}
\label{m15}
\end{eqnarray}
is the momentum resolved current. By means of the equations of motion method, a general expression
for the contour ordered Green function can be found in terms of the AQD and BEC Green's function, $G$ and $g_{\mathbf{k}}$ respectively:
\begin{equation}
G_{0,\mathbf{k}}\left( t,t^{^{\prime }}\right) =\int dt_{1}G\left(
t,t_{1}\right) V_{\mathbf{k}}^{\ast }\left( t_{1}\right) g_{\mathbf{k}%
}\left( t_{1},t^{\prime }\right) .  \label{m16}
\end{equation}
By applying the Langreth rules \cite{jauho2} and using the explicit
expression for the BEC Green's function, i.e. $g_{\mathbf{k}}^{<}\left(
t,t^{^{\prime }}\right) =if_{BE}\left( \varepsilon _{\mathbf{k}}\right)
e^{-i\varepsilon _{\mathbf{k}}\left( t-t^{\prime }\right) },$ and $g_{%
\mathbf{k}}^{r,a}\left( t,t^{^{\prime }}\right) =\\
=\mp i\theta \left( \pm t\mp
t^{^{\prime }}\right) e^{-i\varepsilon _{\mathbf{k}}\left( t-t^{\prime
}\right) }$, where $f_{BE}$ is the Bose distribution function of the
reservoir, one can rewrite the momentum resolved particle current as:

\begin{equation}
 J_{\mathbf{k}}(t)=-\frac{2}{\hbar }%
Im\left\{ V_{\mathbf{k}}\left( t\right) \int_{-\infty }^{t}dt_{1}
V_{\mathbf{k}}^{\ast }\left( t_{1}\right) e^{i\varepsilon _{%
\mathbf{k}}\left( t-t_{1}\right) }\left[ G^{r}\left( t,t_{1}\right)
f_{BE}\left( \varepsilon _{\mathbf{k}}\right) +G^{<}\left( t,t_{1}\right) %
\right] \right\}.
\label{eq:spin-resolved}
\end{equation}

In terms of the self-energies the momentum resolved particle current becomes:
\begin{equation}
J_{\mathbf{k}}(t) =-\frac{2}{\hbar }%
Re\left\{ \int_{-\infty }^{t}dt_{1}\left[
G^{r}\left( t,t_{1}\right) \Sigma_{\mathbf{k}}^{<}\left(
t_{1},t^{^{\prime }}\right) +G^{<}\left( t,t_{1}\right) \Sigma_{%
\mathbf{k}}^{a}\left( t_{1},t^{^{\prime }}\right) \right] \right\} ,
\end{equation}
where the advanced and lesser self-energies of the reservoir are $%
\Sigma_{\mathbf{k}}^{a}\left( t_{1},t^{^{\prime }}\right)=\\
=\left| V_{\mathbf{k}}\right| ^{2}g_{\mathbf{k}}^{a}\left(
t_{1},t^{^{\prime }}\right) $ and $\Sigma_{\mathbf{k}%
}^{<}\left( t_{1},t^{^{\prime }}\right) =\left| V_{\mathbf{k%
}}\right| ^{2}g_{\mathbf{k}}^{<}\left( t_{1},t^{^{\prime }}\right) $.

When the dot level is varied adiabatically in
time, i.e. $\varepsilon (t)=\varepsilon _{0}+\varepsilon_{\omega}\cos (\omega t)$
with $\varepsilon _{\omega}\ll \varepsilon _{0}$, the time dependence of the
tunneling $V_{\mathbf{k}}\left( t\right) $ can be neglected and $|V_{\mathbf{%
k}}|^{2}\approx (\hbar \Omega)^{2}e^{-\left( \sigma
a\mathbf{k}^{2}/2\right) }$, where $\sigma$ and $a$ are the ground state
size of the dot and the BEC in the loosely confined direction, respectively. Let us
note that, since $\sigma $ should be larger than the mean free
distance between the particles $\sigma \gg l$ and $l\sim \xi $
where $\xi $ is the healing length $\xi =\hbar /mc$, it follows
that $\sigma \gg \xi $ and we take it equal to $10^{-6}$m while
 $a$ is taken $10^{-4}m$\cite{aqd3}. In general, for very tightly confined AQD
 the $\mathbf{k}$ dependence of $V_{\mathbf{k}}$ can be neglected.

In this nonequilibrium problem all Green's functions and
self-energies depend explicitly on two time variables, so one
needs to resort to two time Fourier transform and Eq.
(\ref{eq:spin-resolved}) becomes:
\begin{eqnarray}
J_{\mathbf{k}}(t)  &=&-\frac{2}{\hbar }%
Re\left\{ \int \frac{dE_{1}dE_{2}dE_{3}}{\left(
2\pi \right) ^{3}}e^{i\left( E_{3}-E_{1}\right) t}\left[ G^{r}\left(
E_{1},E_{2}\right) \Sigma^{<}\left(
E_{2},E_{3}\right) \right.\right. \nonumber\\
&&\left. \left.+G^{<}\left( E_{1},E_{2}\right)\Sigma_{%
\mathbf{k}}^{a}\left( E_{2},E_{3}\right) \right] \right\},  \label{fin1}
\end{eqnarray}

The lesser and retarded Green function for the dot $G^{<,r}\left( E_{1},E_{2}\right) $ are given by:
\begin{equation}
G^{<}\left( E_{1},E_{2}\right) =\sum_{\mathbf{k}}\int \frac{d\xi _{1}d\xi _{2}}{\left( 2\pi
\right) ^{2}}G^{r}\left( E_{1},\xi _{1}\right) \Sigma_{\mathbf{k}}^{<}\left(
\xi _{1},\xi _{2}\right) G^{a}\left( \xi _{2},E_{2}\right),  \label{gl1}
\end{equation}
\begin{equation}
G^{r}\left( E_{1},E_{2}\right) =\widetilde{G}^{r}\left( E_{1},E_{2}\right)
+\sum_{\mathbf{k}}\int \frac{dE_{3}dE_{4}}{\left( 2\pi \right) ^{2}}\widetilde{G}^{r}\left(
E_{1},E_{3}\right) \Sigma_{\mathbf{k}}^{r}\left( E_{3},E_{4}\right)
\widetilde{G}^{r}\left( E_{4},E_{2}\right) ,  \label{gr1}
\end{equation}
where the zeroth-order dot Green's function in the collisional blockade limit is\\ $\widetilde{G}^{r}\left( t,t_{1}\right)
=-i\theta \left( t-t_{1}\right) e^{-i\int_{t_{1}}^{t}dt^{^{\prime
}}\varepsilon \left( t^{\prime }\right) }=-i\theta \left( t-t_{1}\right)
e^{-i\int_{t_{1}}^{t}dt^{^{\prime }}\left[ \varepsilon _{0}+\varepsilon
_{\omega}\cos \left( \omega t^{\prime }\right) \right] }$
while the reservoir self-energies
$\Sigma^{<}=\sum_{\mathbf{k}}\Sigma_{\mathbf{k}}^{<}$ and
$\Sigma^{r}=\sum_{\mathbf{k}}\Sigma_{\mathbf{k}}^{r}$ are,
respectively:
\begin{eqnarray}
\Sigma^{<}\left( E_{1},E_{2}\right)  &=&\sum_{\mathbf{k}} \left|
V_{\mathbf{k}}\right| ^{2}g_{\mathbf{k}}^{<}\left( E_{1},E_{2}\right)\nonumber\\
=-i\left( 2\pi \right) ^{2}\sum_{\mathbf{k}}f_{BE}\left( \varepsilon _{\mathbf{k}}\right)
\left| V_{\mathbf{k}}\right| ^{2}\delta \left(
E_{1}-\varepsilon _{\mathbf{k}}\right) \delta \left( \varepsilon _{\mathbf{k}%
}-E_{2}\right) , \\
\Sigma^{r}\left( E_{1},E_{2}\right)  &=&\sum_{\mathbf{k}} \left|
V_{\mathbf{k}}\right| ^{2}g_{\mathbf{k}}^{r}\left( E_{1},E_{2}\right)\nonumber\\
=-\left( 2\pi \right)\sum_{\mathbf{k}}\left| V_{\mathbf{k}}\right| ^{2}%
\frac{\delta \left( E_{1}-E_{2}\right) }{E_{1}-\varepsilon _{\mathbf{k}%
}+i0^{+}}=\Lambda(E_1)-i \Gamma(E_1),
\end{eqnarray}
where we have defined $\Lambda(E_1)$ and $\Gamma(E_1)$ as the
principal value and imaginary part of
$1/(E_1-\epsilon_{\mathbf{k}}+i0^+)$. By substituting the
expression of $\tilde{G}^r$ and the self-energy in Eqs.
(\ref{gl1}) and (\ref{gr1}) and taking the limit
$\frac{\varepsilon _{\omega}}{\omega }<<1$, and keeping only terms
linear in $\omega$ we get the following dot Green's functions,
within single photon approximation:
\begin{equation}
G^{r}\left( E_{1},E_{2}\right) =2\pi \left[ A_{1}\delta \left(
E_{1}-E_{2}\right) +B_{1}^{+}\delta \left( E_{1}-E_{2}+\omega \right)
+B_{1}^{-}\delta \left( E_{1}-E_{2}-\omega \right) \right] ,  \label{ret1}
\end{equation}
and
\begin{eqnarray}
G^{<}\left( E_{1},E_{2}\right)  &=&i\Gamma \left( E_{2}\right) \left[
f_{BE}\left( E_{1}\right) A_{1}A_{2}^{\ast }\delta \left( E_{2}-E_{1}\right)
+A_{1}\left( B_{2}^{+}\right) ^{\ast }\delta \left( E_{2}-E_{1}+\omega
\right) +\right.   \nonumber \\
&&\left. A_{1}\left( B_{2}^{-}\right) ^{\ast }\delta \left(
E_{2}-E_{1}-\omega \right) +B_{1}^{+}A_{2}^{\ast }\delta \left(
E_{2}-E_{1}-\omega \right) +\right.   \nonumber \\
&&\left. B_{1}^{-}A_{2}^{\ast }\delta \left(
E_{2}-E_{1}+\omega \right) \right] ,  \label{less1}
\end{eqnarray}
where we have defined:
\begin{equation}
\begin{array}{cc}
A_{1,2}=\frac{\left( J_{0}\left( \frac{\varepsilon _{\omega}}{\omega }\right)
\right) ^{2}}{E_{1}-\varepsilon _{0}-\Lambda(E_{1,2})+i\Gamma \left( E_{1,2}\right) }, &
B_{1,2}^{\pm }=\frac{\omega J_{1}\left( \frac{\varepsilon _{\omega}}{\omega }%
\right) J_{0}\left( \frac{\varepsilon _{\omega}}{\omega }\right) }{\left(
E_{1,2}-\varepsilon _{0}-\Lambda(E_{1,2})+i\Gamma \left( E_{1,2}\right) \right) \left( E_{1,2}\pm
\omega -\varepsilon _{0}-\Lambda(E_{1,2})+i\Gamma \left( E_{1,2}\right) \right) }.
\end{array}
\label{coeff}
\end{equation}
Here $J_{0}\left(\frac{\varepsilon _{\omega}}{\omega } \right) $ and $J_{1}\left(\frac{\varepsilon _{\omega}}{\omega } \right) $ are the $0$-th and $1$-th order Bessel functions, respectively.
By inserting Eqs. (\ref{ret1})-(\ref{less1}) in Eq. (\ref{fin1}), we can numerically evaluate
both the energy absorption
rate (EAR) and the particle current.  The behavior of the EAR and the particle current, within single photon approximation, are shown in Figs. 1 and
2.

\begin{figure}[h]
\resizebox{0.75\columnwidth}{!}{\includegraphics{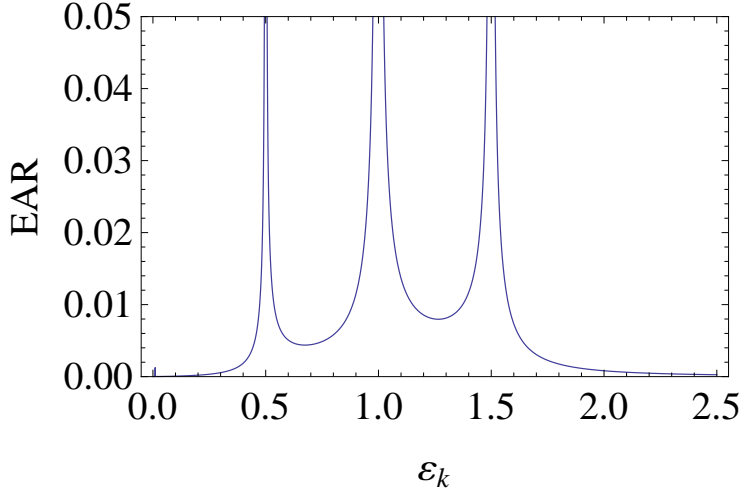}}
\caption{Energy absorption rate as a function of $\epsilon_k$ for the following values of parameters (in unit of the quantum dot energy level): $\epsilon_0=1.0$, $\Omega=0.1$, $\epsilon_\omega=0.06$, $\omega=0.5$, $T=1$. The time is fixed at $t=1.0$ in units of the inverse of $\omega$.} \label{fig:1}
\end{figure}
As shown, the EAR spectrum shows a peak at $\epsilon_0$ and two side peaks corresponding to the absorption and emission of a photon with energy $\omega$.
\begin{figure}[h]
\resizebox{0.75\columnwidth}{!}{\includegraphics{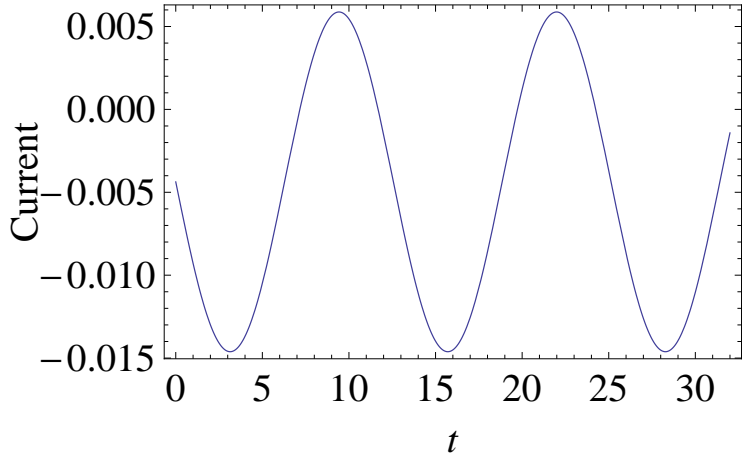}}
\caption{Particle current as a function of time (in units of $\omega^{-1}$) for the following values of parameters (in unit of the quantum dot energy level): $\epsilon_0=1.0$, $\Omega=0.1$, $\epsilon_\omega=0.06$, $\omega=0.5$, $T=1$.} \label{fig:2}
\end{figure}

The current induced by the quantum dot modulation follows the harmonic variation of the perturbation and its amplitude and mean value get modified as a result of single-photon absorption and emission processes.

\section{Conclusions}
\label{sec:2}
The nonequilibrium properties of an atomic quantum dot (AQD)
coupled to a Bose-Einstein condensate (BEC) were studied within the Keldysh-Green's
function formalism when the AQD level is varied harmonically in
time. We have analyzed both the current and the AQD energy absorption spectrum in the full nonequilibrium situation, within a single-photon approximation and shown that it is possible to achieve an efficient way of transferring particles from the BEC to the AQD in a range of parameters interesting for current experiments with cold atoms. Both quantities are actually a subject of active investigation theoretically as well as experimentally in order to probe and manipulate such systems.

Indeed the coherent tunneling of particles from the BEC to the AQD could be employed to extract atoms on demand from the quantum reservoir, thus realizing a quantum tweezer\cite{morigi1}. This high degree of control is needed in several protocols for quantum information processing with neutral atoms, in particular at the initialization stage of a quantum register \cite{qreg1}. Further applications can be envisaged, which run from the optimal control of atoms with microwave potentials for the implementation of quantum gates on an atom chip \cite{treut1} to the realization of an efficient procedure to filter out from an optical lattice a preselected number of atoms per site \cite{niko1}. In this way lattices with a desired site occupation could be engineered.

On the other hand, the energy absorption rate could be employed as a spectroscopic tool to probe several properties of the bosonic system under study, and in particular the current autocorrelation function, in analogy with a recent proposal by Giamarchi and co-workers \cite{giamarchi}. That allows one to directly probe the frequency dependent conductivity of the system.

\end{document}